\documentclass[aps,prl,reprint,groupedaddress]{revtex4-1}

\usepackage{graphicx}
\usepackage{dcolumn}
\usepackage{bm}
\usepackage{amsbsy}
\usepackage{graphicx}
\usepackage{hyperref}
\usepackage{graphics}
\usepackage{mathrsfs}

\usepackage[utf8]{inputenc}

\newcommand{\be}{\begin{equation}} \newcommand{\ee}{\end{equation}}
\newcommand{\bea}{\begin{eqnarray}} \newcommand{\eea}{\end{eqnarray}}

\newcommand{\re}[1]{(\ref{#1})}

\newcommand{\fig}[1]{figure \ref{#1}}

\newcommand{\para}{\paragraph}

\newcommand{\LCDM}{$\Lambda$CDM\ }

\newcommand{\rmd}{\mathrm{d}}

\newcommand{\ie}{i.e.\ }

\newcommand{\adot}{\dot{a}}

\newcommand{\OKn}{\Omega_{K0}}

\renewcommand{\l}{\mathrm{l}}
\newcommand{\s}{\mathrm{s}}
\newcommand{\dls}{d_\mathrm{ls}}
\newcommand{\dl}{d_\mathrm{l}}
\newcommand{\ds}{d_\mathrm{s}}

\begin{document}

\onecolumngrid
\begin{flushright}
  HIP-2019-16/TH \\
  IFT-UAM/CSIC-19-70
\end{flushright}
\twocolumngrid

\title{Model-independent determination of $H_0$ and $\Omega_{K0}$ \\ from strong lensing and type Ia supernovae}

\author{Thomas Collett}
\affiliation{Institute of Cosmology and Gravitation, University of Portsmouth, Dennis Sciama Building, Burnaby Road, Portsmouth, PO1 3FX, United Kingdom}

\author{Francesco Montanari}
\affiliation{Instituto de F\'isica Te\'orica IFT-UAM/CSIC, Universidad
  Aut\'onoma de Madrid, Cantoblanco 28049 Madrid, Spain}

\author{Syksy R\"{a}s\"{a}nen}
\affiliation{University of Helsinki, Department of Physics and Helsinki Institute of Physics, P.O. Box 64, FIN-00014 University of Helsinki, Finland}


\begin{abstract}

We present the first determination of the Hubble constant $H_0$ from strong lensing time delay data and type Ia supernova luminosity distances that is independent of the cosmological model. We also determine the spatial curvature model-independently. We assume that light propagation over long distances is described by the FLRW metric and geometrical optics holds, but make no assumption about the contents of the Universe or the theory of gravity on cosmological scales. We find $H_0=75.7^{+4.5}_{-4.4}$ km/s/Mpc and $\Omega_{K0}=0.12^{+0.27}_{-0.25}$. This is a 6\% determination of $H_0$. A weak prior from the cosmic microwave background on the distance to the last scattering surface improves this to $H_0=76.8^{+4.2}_{-3.8}$ km/s/Mpc and $\Omega_{K0}=0.18^{+0.25}_{-0.18}$.  Assuming zero spatial curvature, we get $H_0=74.2^{+3.0}_{-2.9}$ km/s/Mpc, a precision of $4\%$. The measurements also provide a consistency test of the FLRW metric: we find no evidence against it.

\end{abstract}



\maketitle

\setcounter{tocdepth}{2}

\setcounter{secnumdepth}{3}

\section{Introduction} \label{sec:intro}

\para{Determining the Hubble constant and spatial curvature.}

The value of the Hubble constant $H_0$ has emerged as the strongest point of tension between predictions of the \LCDM model of cosmology and observations. Fitting the \LCDM model to cosmic microwave (CMB) data of the Planck satellite gives $H_0=67.4\pm0.5$ km/s/Mpc \cite{Planck:cosmo}. (Our error bars are 68\% limits, while ranges and inequalities are 95\% limits.) On the other hand, distance ladder measurements of local type Ia supernovae (SNe) that are only weakly dependent \cite{Feeney:2017sgx} on the cosmological model give $H_0=74.03\pm1.42$ km/s/Mpc, over 4$\sigma$ away from the Planck result \cite{Riess:2019}. Several studies have not found any systematics that could explain the difference \cite{Efstathiou:2013via, Riess:2016jrr, Cardona:2016ems, Zhang:2017aqn, Follin:2017ljs, Riess:2018uxu, Riess:2018byc, Feeney:2017sgx}.

Independent determinations of $H_0$ can provide clues about the origin of the discrepancy. The time delay between strongly lensed images of time variable sources is inversely proportional to the Hubble constant \cite{refsdal64}. Time delay measurements therefore provide the time scale for $H_0^{-1}$ directly, without the need for a distance ladder, and time delays depend only on late universe physics, unlike the CMB and baryon acoustic oscillations (BAO). In order to determine $H_0$ from time delays, the dimensionless distances between the observer, lens and source are also needed. In all of the determinations so far, this has been done by assuming some cosmological model. Recently the H0LiCOW project \cite{h0licow1} has derived $H_0 = 72.0^{+2.3}_{-2.6}$ km/s/Mpc \cite{h0licow9} in the spatially flat $\Lambda$CDM model using a sample of four lenses \cite{h0licow9,h0licow4,suyu2010,suyu2013}. The inferred value of $H_0$ is strongly model-dependent. The first three H0LiCOW lenses yielded $H_0 = 71.9^{+2.4}_{-3.0}$ km/s/Mpc assuming the spatially flat $\Lambda$CDM model, but this changes to $79.1^{+9.3}_{-8.7}$ km/s/Mpc if the dark energy equation of state is not fixed to $-1$ \cite{h0licow5}.

The model-dependence can be reduced by determining the distances from the observer to the lens and source directly from observations of type Ia SNe. However, the relation of these distances to the distance from the lens to source cannot be determined directly from the observations. In Friedmann--Lema\^{\i}tre--Robertson--Walker (FLRW) universes it depends on the spatial curvature. Turning this around, we can use combined observations of time delays and SNe distances to determine not only $H_0$ but also the spatial curvature model-independently, assuming that the universe is described by the FLRW metric. Furthermore, by observing two or more lens-source pairs and comparing the inferred values of the spatial curvature, we in principle have a consistency test for the FLRW metric. Failure of the FLRW approximation could be related to extra dimensions \cite{Ferrer:2005hr, *Ferrer:2008fp, *Ferrer:2009pq}, violation of statistical homogeneity and isotropy \cite{Enqvist:2007vb, *February:2009pv, *Bolejko:2011jc, *Redlich:2014gga}, or the effect of deviation from exact homogeneity and isotropy on the average expansion rate, \ie backreaction \cite{Rasanen:2008be, *Rasanen:2009uw, *Buchert:2011sx, *Boehm:2013}.

For strong lensing, a consistency condition based on image deformation has been proposed and implemented \cite{Rasanen:2014, Xia:2016}. The method was proposed to be extended to time delays in \cite{Rasanen:2014} and estimates have been done on simulated data \cite{Liao:2017, Denissenya:2018, Liao:2019, Li:2019rns, Qi:2018atg}. We will now for the first time apply it to real data. A consistency condition based on comparing distance and expansion rate has also been implemented \cite{Clarkson:2007, Mortsell:2011yk, Heavens:2014, Yu:2016gmd, Wei:2016xti, Li:2016wjm, Shafieloo:2009hi, Sapone:2014nna, LHuillier:2016mtc, Montanari:2017, Wang:2017lri, Cai:2015pia, LHuillier:2016mtc, Shafieloo:2018gin}, and proposed for luminosity distance and parallax distance \cite{Rasanen:2013}.

\section{Theoretical aspects} \label{sec:cons}

\para{Distances.}

If space is exactly homogeneous and isotropic, spacetime is described by the FLRW metric
\bea \label{metric}
  \rmd s^2 = - \rmd t^2 + \frac{a(t)^2}{1 - K r^2} \rmd r^2 + a(t)^2 r^2 \rmd\Omega^2 \ ,
\eea

\noindent where $K$ is a constant related to the spatial curvature. The Hubble parameter is $H\equiv\adot/a$, and its present value is denoted by $H_0$. Let $D_A(z_\l,z_\s)$ be the angular diameter distance of a source at redshift $z_\s$ as seen at redshift $z_\l$. From \re{metric} we find, assuming that geometrical optics holds, $t(z)$ is monotonic, that the dimensionless distance $d(z_\l,z_\s)\equiv (1+z_\s) H_0 D_A(z_\l,z_\s)$ (which is independent of $H_0$) is
\bea \label{bidt}
  d(z_\l, z_\s)
  &=& \frac{1}{\sqrt{\OKn}} \sinh\left( \sqrt{\OKn}\int_{z_\l}^{z_\s} \frac{H_0}{H(z)} \rmd z \right) \ ,
\eea

\noindent where $\OKn\equiv -K/H_0^2$. For $\OKn=0$ the expression reduces to a linear function of the integral in the argument, and for $\OKn<0$ it becomes a sine function of the argument. We denote $d(z)\equiv d(0,z)$.

\para{Distance and spatial curvature.}

We assume that $d(z)$ is monotonic. Using \re{bidt}, $\dls\equiv d(z_\l,z_\s)$ can then be written in terms of $\dl\equiv d(z_\l)$ and $\ds\equiv d(z_\s)$ as \cite{Peebles:1993, Bernstein:2005, Rasanen:2014}
\bea \label{sum}
  \dls &=& \ds \sqrt{1 + \OKn \dl^2} - \dl \sqrt{1 + \OKn \ds^2} \ .
\eea
As noted in \cite{Rasanen:2014}, we can solve for the spatial curvature in \re{sum} to get a consistency condition for the FLRW metric:
\bea \label{k}
  k_S &=& - \frac{\dl^4 + \ds^4 + \dls^4 - 2 \dl^2 \ds^2 - 2 \dl^2 \dls^2 - 2 \ds^2 \dls^2}{4 \dl^2 \ds^2 \dls^2} \ ,
\eea
where $k_S\equiv-\OKn$. If the combination of distances \re{k} is observationally found not to be equal for any two pairs of $(z_\l, z_\s)$, the FLRW metric is ruled out.

\para{Distance and time delay.}

In strong lensing, light propagating from the source splits into several bundles to form multiple images. The difference $\Delta t_{12}\equiv t_2-t_1$ in the arrival times $t_1$ and $t_2$ of two images labeled 1 and 2 at angular coordinates $\theta_1$ and $\theta_2$ on the sky is \cite{Narayan:1996ba}
\bea \label{time}
  \Delta t_{12}(\theta_1,\theta_2) = H_0^{-1} \frac{\dl \ds}{\dls} f(\theta_1,\theta_2) = D_{\Delta t} f(\theta_1,\theta_2) \ ,
\eea
where $f(\theta_1,\theta_2)$ depends on the structure of the lens and the second equality defines the time delay distance $D_{\Delta t} f(\theta_1,\theta_2)$. Given observations of $\Delta t_{12}$, $\dl$ and $\ds$ and a model for the lens, we can determine $H_0$ and $\OKn$ from \re{sum} and \re{time}.

\section{Observations} \label{sec:det}

\para{Supernova data.}

The Pantheon compilation \cite{Scolnic:2017caz} provides distance moduli $\mu_i$ up to an absolute magnitude $M$ for 1048 SNe that we collect into the vector $\hat{{\boldsymbol X}}$, where $X_i = \mu_i + M$ and the hat denotes Pantheon data, and its covariance matrix ${\boldsymbol C}$ (including both statistical and systematic uncertainties). We rely on the reduced $\hat{{\boldsymbol X}}$ data given the fit to light-curve parameters and the fit to their coefficients entering in the ${\boldsymbol X}$ estimator with simultaneous bias corrections performed by the Pantheon collaboration. The highest redshift in the compilation is 2.3. Luminosity distances $D_L$ can be inferred with arbitrary overall normalisation maximizing the likelihood $\mathcal{L}$ defined by $-2\log\mathcal{L} = \left(\hat{{\boldsymbol X}} - {\boldsymbol X}\right)^T {\boldsymbol C}^{-1} \left(\hat{{\boldsymbol X}} - {\boldsymbol X}\right)$. The model vector ${\boldsymbol X}$ is determined by  $\mu_i = 5 \log_{10}\left[D_L({\boldsymbol \theta}, z_i)/(10\ {\rm pc})\right] = 5 \log_{10}d_L({\boldsymbol \theta}, z_i) + M_{H_0}$, where ${\boldsymbol \theta}$ denotes the model parameters and $z_i$ the SN redshift. Given the degeneracy with the absolute magnitude $M$, the value of $M_{H_0}=-5 \log_{10}(10\ {\rm pc}\ H_0)$ is arbitrary and we fix it to $M_{H0} \approx 43.2$ (corresponding to an nonphysical $H_0=70$~km/s/Mpc). Note that in any spacetime for any metric theory of gravity  $d_L\equiv H_0 D_L=(1+z) d$ holds true \cite{Etherington:1933, Ellis:1971pg}.

The data was preprocessed by the Pantheon team by fitting coefficients that relate the light-curve parameters to the distance modulus, and simultaneously correcting related biases. This analysis is  model-dependent, because bias corrections assume dark energy with a constant equation of state \cite{Kessler:2016uwi, Scolnic:2017caz}. The dependence on the cosmological model is marginal for changes in the reference cosmology within typical statistical uncertainties. Furthermore, studies using the JLA SN dataset \cite{Betoule:2014frx}, dependent on bias corrections done assuming the \LCDM model (and specific values of light-curve parameters), have found that the model-dependence of the light-curve parameters is weak \cite{Marriner:2011mf, Wei:2016xti, Li:2016wjm, Hauret:2018, Montanari:2017}. Therefore, this model-dependence is likely a subdominant source of bias. There are also significant differences between light curve fitters (for discussion of these and other systematics, see \cite{Nadathur:2010, Bengochea:2010it, *Li:2010du, *March:2011xa, *Lago:2011pk, *Giostri:2012ek, *Kessler:2012gn, *Nielsen:2015pga, *Shariff:2015yoa, *Rubin:2016iqe, *Dam:2017xqs, *Tutusaus:2017ibk}), but they are also likely a subdominant source of errors.

\para{Strong lensing data.}

There are currently four strong lensing systems with accurately modelled time delay distances: B1608+656 \cite{suyu2010}, RXJ1131--1231 \cite{suyu2013}, HE0435--1223 \cite{h0licow4} and SDSSJ1206+4332 \cite{h0licow9}. The highest source redshift is 1.789, well below the maximum redshift of the Pantheon SNe. For 1608, 1131 and 0435 we use the skewed log-normal approximations to the likelihood functions derived in \cite{h0licow5}. For 1206 we approximate the likelihood as skewed log-normal with $\mu_D = 7.8817$, $\sigma_D = 0.2016$, $\lambda_D = 3127.4$, consistent with figure 12 of \cite{h0licow9}. These parameters are defined in equation 3 of \cite{h0licow5}. We verified that using the full likelihood \cite{h0licow9} leads to negligible changes in our results. The time delay distances are weakly dependent on the assumed cosmology, which enters into the line-of-sight lensing \cite{h0licow3, h0licow4}. This was investigated in \cite{h0licow4}, which found that the choice of cosmological parameters impacted the line-of-sight lensing level at the $\mathcal{O}(0.5\%)$ level for 0435. On average lens lines-of-sight are expected to be almost the same as random lines through the Universe \cite{collettcunnington2016}, so the effect of changing the cosmology is expected to be small. However, lenses live in locally over dense regions \cite{fassnacht2011}. This effect is calibrated using the Millennium Simulation \cite{springel2005}; changing the cosmology used in this calibration could plausibly change the inferred time delay distances by 1\%, though it is worth noting that a more direct calibration using weak lensing \cite{h0licow8} for 0435 is in agreement with the Millennium Simulation method used in \cite{h0licow4}.

We also use constraints from the compound gravitational lens SDSSJ0946+1006 \cite{gavazzi2008}. Here the presence of two sources $s1$ and $s2$ lensed by the same foreground mass enables a precise constraint on the cosmological scaling factor $\beta = \frac{d_{l,s1} d_{s2}}{d_{s1} d_{l,s2}}$ \cite{collettauger2014}. This ratio has no dependence on $H_0$, but it is sensitive to spatial curvature. We neglect the uncertainty of the redshift $z_{s2}$ of the second source, since for the fiducial cosmology it produces changes in $\beta$ that are less than half the measurement error, and take the redshift to be at the peak of the photometric redshift probability from \cite{collettauger2014}, \ie $z_{s2}=2.3$. This is at the end of the range of the Pantheon SN dataset.

\section{Datafit and results} \label{sec:res}

\para{Fitting function for $d(z)$.}

We obtain $d(z)$, $H_0$ and $\OKn$ model-independently by fitting the SN and time delay data simultaneously using a Markov chain Monte Carlo sampler \cite{emcee}. We model the function $d(z)$ with a polynomial. By fitting to mock Pantheon data for different \LCDM models (which include spatial curvature) and performing an out-of-sample error analysis based on real data, we have found that a fourth order polynomial is versatile enough to fit current data, while higher order polynomials do not improve the goodness-of-fit taking into account the number of free parameters. We also find that the typical offset of the mean from the real underlying value is less than $\sim 10\%$ of the statistical uncertainty. This agrees with the analyses performed for Union2.1 \cite{Rasanen:2014} and JLA \cite{Montanari:2017} data. Splines, rational functions and B\'ezier curves were also considered in \cite{Rasanen:2014, Montanari:2017}, finding no improvement over polynomials. As $d(0)=0$ and $d'(0)=1$, the fourth order polynomial has three free parameters. The absolute magnitude $M$ enters in the SN likelihood as a nuisance parameter. In addition, $\dls$ given by \re{sum} involves the constant $\OKn$ and the time delays involve $H_0$, giving six parameters in total. For comparison, we also fit the \LCDM model (with and without spatial curvature) to the data.

\para{Priors.}

As flat priors, we take $0 < H_0\ {\rm [km/s/Mpc]} < 150$, $-2 < \OKn < 2$ and $-25 < M < -15$. When fitting polynomials, the order $i$ coefficients $c_i$ are varied within $-10 < c_i < 10$. When fitting the \LCDM model we take $0 < \Omega_{\Lambda0} < 1.5$.

We also consider the following more informative priors. For $\OKn$, we use a prior obtained from the model-independent value $D_A(0,1090)=12.8\pm0.07$ Mpc from the CMB \cite{Vonlanthen:2010cd, *Audren:2012wb, *Audren:2013nwa} and the conservative bound $H_0>60$ km/s/Mpc, which combine to give $d(1090)>2.8$. From \re{sum} this translates into $\OKn>-0.1$. If $\OKn$ were more negative (corresponding to the spatial curvature being more positive) than indicated by this limit, the universe would be too small to contain the last scattering surface. We also consider the flat case $\OKn=0$. For $H_0$, we consider the prior $H_0=74.03\pm1.42$ km/s/Mpc from local SNe \cite{Riess:2019}. When fitting the \LCDM model we always impose $\Omega_{m0} > 0$ (this restriction is relevant when including time delay data only).

\begin{table*}[t!]
\begin{center}
\begin{tabular}{|c|ccc|c|}
\hline
Model & $H_0$ [km/s/Mpc] & $\Delta H_0/H_0$  & $\OKn$  & Restrictions \\
\hline
Polynomial  & $75.7^{+4.5}_{-4.4}$ & 6\% & $0.12^{+0.27}_{-0.25}$ & None \\
Polynomial  & $76.8^{+4.2}_{-3.8}$ & 5\% & $[-0.08,0.73]$ & $\OKn>-0.1$ \\
Polynomial  & $74.2^{+3.0}_{-2.9}$ & 4\% & -  & $\OKn=0$ \\
Polynomial  & $74.2^{+1.3}_{-1.3}$ & 2\% & $0.05^{+0.18}_{-0.17}$  & $H_0=74.03\pm1.42$ km/s/Mpc \\
\hline
\LCDM  & $72.9^{+2.4}_{-2.4}$ & 3\% & $0.00^{+0.16}_{-0.16}$   & None \\
\LCDM  & $73.4^{+2.3}_{-2.4}$ & 3\% & $[-0.09,0.36]$   & $\OKn>-0.1$ \\
\LCDM  & $73.0^{+2.1}_{-2.3}$ & 3\% & -  & $\OKn=0$  \\
\LCDM  & $73.8^{+1.2}_{-1.2}$ & 2\% & $0.02^{+0.15}_{-0.14}$  & $H_0=74.03\pm1.42$ km/s/Mpc \\
\LCDM  & -  & - & $-0.06^{+0.18}_{-0.17}$  & SN data only \\
\LCDM  & $73.5^{+2.9}_{-2.9}$ & 4\% & $0.25^{+0.32}_{-0.32}$  & time delay data only \\
\LCDM  & $72.3^{+2.3}_{-2.5}$ & 4\% & -  & time delay data only, $\OKn=0$ \\
\hline
\end{tabular}
\end{center}
\caption{Results for $H_0$ and $\OKn$ for the polynomial fit and the \LCDM model with various choices of priors and data. In the case of the prior $\OKn>-0.1$ we show the 95\% range for $\OKn$, as the distribution is far from Gaussian.}
\label{tab:res}
\end{table*}

\para{Values of $H_0$ and $\OKn$.}

\begin{figure}[t]
\scalebox{0.5}{\includegraphics[angle=0, clip=true, trim=0.0cm 0.0cm 0.0cm 0.0cm]{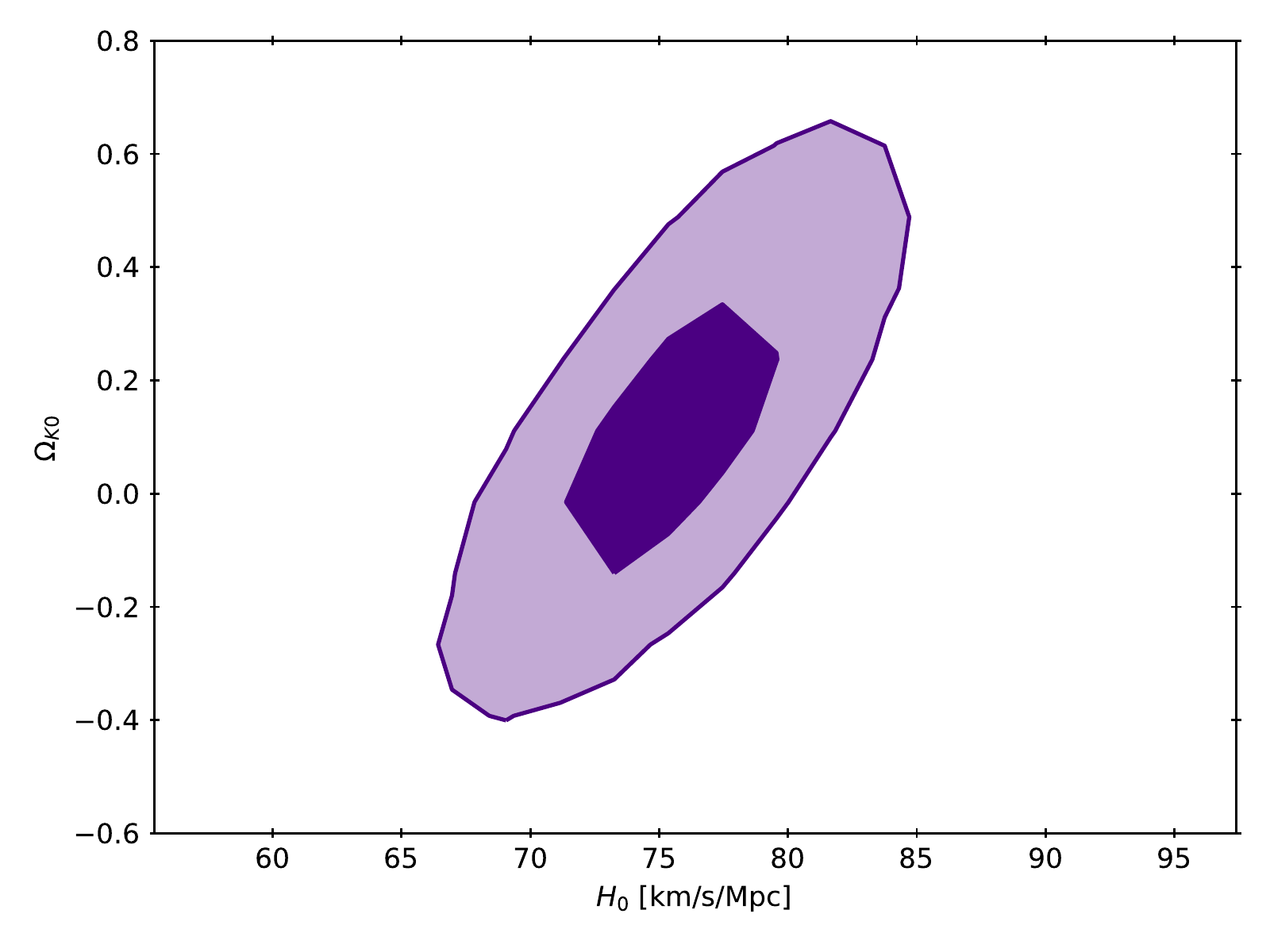}}
\caption{The 68\% and 95\% confidence contours of $H_0$ and $\OKn$ in the polynomial case with flat priors.}
\label{fig:P}
\end{figure}

We marginalise over the three polynomial coefficients (in the \LCDM case, over $\Omega_{\Lambda0}$) and the SN absolute magnitude to obtain the probability distributions for $H_0$ and $\OKn$. The results are given in table \ref{tab:res}. Comparable values of maximum likelihood suggest similar performance of the model-independent and \LCDM fits for all cases investigated here. The posteriors are close to Gaussian except when we have the weak CMB prior $\OKn>-0.1$, so for that case we show the 95\% range rather than the 68\% error bars. Compared to the case when the spatial curvature is free to vary, the prior $\OKn>-0.1$ reduces the error bars on $H_0$ by 20\%. Assuming spatial flatness reduces the error bars by 30\%. These determinations of $H_0$ have a precision of 6\%, 5\% and 4\%, respectively. For comparison, the precision from local SNe is 2\% \cite{Riess:2019}. The limits on the spatial curvature are $-0.37 < \OKn < 0.70$ (flat priors), $-0.08 < \OKn < 0.73$ (CMB prior $\OKn > -0.1$) and $-0.28 < \OKn < 0.43$ (local SN prior on $H_0$). There is no evidence for spatial curvature.

In the case with flat priors, these model-independent error bars for both $H_0$ and $\OKn$ are larger than in the \LCDM case by a factor of 2. Even with the CMB prior on $\OKn$, the error bars on $H_0$ grow by 70\%. The reason is that in the model-independent case, the SN data contain no information about spatial curvature, unlike in the \LCDM case. With the prior on $H_0$, the errors on $\OKn$ grow by only 10\%.

In \fig{fig:P} we show the 2D marginalised contours on the $H_0-\OKn$ plane in the model-independent case with flat priors. The CMB prior $\OKn>-0.1$ makes the probability distribution of $\OKn$ highly non-Gaussian due to the top-hat cut, but otherwise the probability contours do not change much. The 2D plot with the CMB prior would look like a truncated version of the plot with the flat prior. As $H_0$ and $\OKn$ are positively correlated, the prior on $\OKn$ slightly increases the value of $H_0$, but the shift is much smaller than the error bars.

Computing the model-independent fits with only one single-source lens system at a time provides an estimate of the distance sum rule \re{k} for different $(z_\l , z_\s)$ pairs and hence a consistency check for the FLRW metric. However, we find that when we have both $H_0$ and $\OKn$ as free parameters, the current data has no constraining power. Comparing model-independent fits where we consider either all single-source systems or the double-source system alone still provides a consistency test of the FLRW metric. Even taken together, the constraints on spatial curvature from single-lens systems are very weak, and we find no evidence against the FLRW metric. This test also shows that including the double lens system in the analysis with all the systems is important, as it drives the constraints on $\OKn$, which in turn helps to improve the determination of $H_0$.

Concentrating on the \LCDM model, the $H_0$ error bars fall by 17\% when we add the SN data to the time delay data. The SN data contains no information about $H_0$, but it helps to constrain the vacuum energy and the spatial curvature. The constraint on the spatial curvature comes mostly from the SN data. Compared to the case with SN data only, the errors on $\OKn$ decrease only 9\%, but compared to the case with time delay data only, they drop by 50\%. Although the mean values of $H_0$ determined model-independently are larger than in the \LCDM case, the results are well consistent within 1$\sigma$ across choices of datasets and priors.

\section{Conclusions} \label{sec:conc}

\para{Results and comparison to previous work.}

The model-independent values for $H_0$ from SN and time delay data are in good agreement with the determination from local SNe. The mean value of $H_0$ is considerably higher than the result from fitting the \LCDM model to the Planck CMB data, from 6.8 to 9.4 km/s/Mpc, depending on the priors, but the difference is always less than 3$\sigma$. These findings are consistent with previous model-dependent determinations of $H_0$ from H0LiCOW data \cite{h0licow9}.

The SN distances and time delays involve only late universe physics, whereas the CMB and BAO are dependent on early universe physics. It has been suggested that the discrepancy of $H_0$ determined from local SNe or the CMB might be due to early universe physics beyond the \LCDM model, in particular a smaller sound horizon \cite{Bernal:2016gxb, Riess:2019}. This is supported by analyses combining BAO and other data \cite{Percival:2009, Aubourg:2014, Alam:2016, Addison:2017, Lin:2017, Macaulay:2018fxi, Lemos:2018smw}. However, extrapolation of the Hubble parameter $H(z)$ determined from cosmic clocks down to $z=0$ gives a small value of $H_0$ more consistent with the CMB data, though the difference does not seem to be significant given current error bars \cite{Busti:2014dua, Verde:2014qea, Busti:2014aoa, Montanari:2017, Wang:2017lri} (see also the combination of SN and cosmic clock data in \cite{Gomez-Valent:2018hwc, *Gomez-Valent:2019lny}, with smaller errors). If the difference persists, this could rather point to a distinction between determinations of $H_0$ based directly on the expansion rate (radial BAO mode and cosmic clocks) and those derived from distances (SNe and time delays), although it is not clear how the CMB would fit this pattern.

The model-independent value for the spatial curvature is determined with an error of $\Delta\OKn=0.2\dots0.3$, the precise value depending on the priors. This is two orders of magnitude worse than the model-dependent \LCDM limit from Planck CMB plus BAO data, $\OKn=0.0007\pm0.0019$ \cite{Planck:cosmo}, driven by the sensitivity of the angular diameter distance to the spatial curvature at large redshifts. However, we have made no assumptions about the matter content or theory of gravity on cosmological scales, only the validity of the FLRW metric and geometrical optics. We also tested the FLRW consistency condition from the distance sum rule introduced in \cite{Rasanen:2014}. We find no evidence of inconsistency of the FLRW metric, but the constraining power of the current time delay data is quite weak.

In the case when we do not impose informative priors, the constraint on $\OKn$ is better by a factor of 2 compared to the value from strong lensing image deformation \cite{Rasanen:2014, Xia:2016} and roughly the same as the best determinations based on comparing SN distances and cosmic clocks \cite{Mortsell:2011yk, Heavens:2014, Yu:2016gmd, Wei:2016xti, Li:2016wjm, Montanari:2017, Wang:2017lri}. In the present case, systematics related to lens modelling are better under control than in previous analyses.

\para{Forecast.}

Let us estimate the expected improvement from upcoming observations, roughly taking as the reference 10 years of LSST \footnote{\url{https://www.lsst.org/}} observations. We consider $10^5$ type Ia SNe logarithmically distributed over the Pantheon redshift range and take fractional errors on the distance modulus to be 0.5\%, roughly the mean value in the JLA and Pantheon datasets, as the SN data is already limited by systematics. For the time delay data, we consider 400 systems \cite{Liao:2014cka} with both the lens and the source in the Pantheon redshift range (the redshift distribution is determined by random draw from \cite{OM2010}), with 7\% fractional errors on the time-delay distance \cite{Shajib:2017omw}. We assume a \LCDM cosmology with $\Omega_{m0}=0.30$, $\Omega_{\Lambda0}=0.69$, $\Omega_{K0}=0.01$ and $H_0=70$ km/s/Mpc. The larger dataset requires using a fifth order polynomial to fit $d(z)$. Compared to the present data, the error bars on both $H_0$ and $\Omega_{K0}$ shrink by a factor of 8: $H_0$ is determined with a precision of 0.9\%, and the error on the spatial curvature is $\Delta\Omega_{K0}=0.03$. This is consistent, within the different assumptions about the systems observed, with previous forecasts based on a combination of strong lensing image deformation and/or time delays with SN data \cite{Rasanen:2014, Liao:2017, Denissenya:2018, Qi:2018atg}, as well using gravitational waves and their electromagnetic counterparts to measure both time delay and luminosity distance from the same systems \cite{Liao:2019, Li:2019rns}, or combining distances from gravitational waves with cosmic clock data \cite{Wei:2018}. The large number of lens systems will also allow to fit the distance sum rule \re{k} as a function of redshift, providing a stronger null test of the FLRW metric.

\para{Acknowledgments.}

We thank Simon Birrer for useful comments and help with H0LiCOW likelihoods and Dan Scolnic for providing a version of the Pantheon catalog before public release and clarifying lightcurve bias corrections. We acknowledge use of the Alcyone (HIP Helsinki) and Hydra (IFT Madrid) clusters. FM is supported by the Research Project FPA2015-68048-C3-3-P [MINECO-FEDER] and the Centro de Excelencia Severo Ochoa Program SEV-2016-0597.

\bibliography{time}

\end{document}